\documentclass[aps,prl,reprint,superscriptaddress]{revtex4-1}

\usepackage{amsmath}
\usepackage{amssymb}
\usepackage{CJK}
\usepackage{graphicx}
\usepackage{dcolumn}
\usepackage{bm}
\usepackage{hyperref}
\usepackage[dvipsnames]{xcolor}
\usepackage[tight]{subfigure} 

\newcommand{\N}{\bm{N}}
\newcommand{\M}{\bm{M}}
\newcommand{\x}{\bm{x}}

\begin{document}
\begin{CJK*}{UTF8}{}

\title{Heterobilayer moir\'e magnets: moir\'e skyrmions and \\ the commensurate-incommensurate transition}

\CJKfamily{gbsn}

\author{Kasra Hejazi}
\thanks{These two authors contributed equally.}
\affiliation{Department of Physics, University of California Santa Barbara, Santa Barbara, California, 93106-4030, USA}
\author{Zhu-Xi Luo (罗竹悉)}
\thanks{These two authors contributed equally.}
\affiliation{Kavli Institute for Theoretical Physics, University of California, Santa Barbara, CA 93106-4030, USA}
\author{Leon Balents}
\affiliation{Kavli Institute for Theoretical Physics, University of California, Santa Barbara, CA 93106-4030, USA}
\affiliation{Canadian Institute for Advanced Research, Toronto, Ontario, Canada}

\date{\today}

\begin{abstract}
We study untwisted heterobilayers of ferromagnetic and antiferromagnetic van der Waals materials, with in particular a Dzyaloshinskii-Moriya interaction in the ferromagnetic layer. A continuum low energy field theory is utilized to study such systems. We develop a phase diagram as a function of the strength of inter-layer exchange and Dzyaloshinskii-Moriya interactions, combining perturbative and strong coupling analyses with numerical simulations using Landau-Lifshitz-Gilbert equations. Various moir\'e-periodic commensurate phases are found, and the commensurate-incommensurate transition is discussed. Among the commensurate phases, we observe an interesting skyrmion lattice phase wherein each moir\'e unit cell hosts one skyrmion.
\end{abstract}

\maketitle
\end{CJK*}

Magnetic skyrmions are long-lived, topologically protected spin-textures that were predicted to exist in chiral magnets \cite{bogdanov1989thermodynamically,bogdanov1994thermodynamically,roessler2006spontaneous}, and experimentally observed in cubic, non-centrosymmetric materials such as MnSi \cite{muhlbauer2009skyrmion}, Fe$_{1- x}$Co$_x$Si \cite{munzer2010skyrmion,yu2010real} and FeGe \cite{yu2011near}. Non-collinear spin configurations occur in these materials due to the antisymmetric Dzyaloshinskii-Moriya (DM) interactions \cite{dzyaloshinsky1958thermodynamic,moriya1960anisotropic}. In three-dimensional bulk magnets, skyrmion lattices are stabilized by thermal fluctuations above the helical state \cite{muhlbauer2009skyrmion}. Interestingly, in the thin-film, two-dimensional limit, the skyrmion lattice is stable over a wide range of the phase diagram \cite{do2009skyrmions, han2010skyrmion, li2011general}. 

A new class of two-dimensional crystals, magnetic van der Walls (vdW) materials, has opened up numerous possibilities for both theoretical and experimental physics \cite{park2016opportunities,burch2018magnetism}. All three fundamental spin Hamiltonians have been reported in these materials: the two-dimensional Heisenberg, Ising, and XY models  \cite{joy1992magnetism,wildes2017magnetic}. Furthermore, novel quantum phases are expected to appear in the heterostructures of these materials \cite{hejazi2020noncollinear}. 
Recently, skyrmion crystals have been observed in the single-layered ferromagnetic vdW material Fe$_3$GeTe$_2$ \cite{wang2019direct,park2019neltype,ding2019observation}, predicted to exist in CrI$_3$ \cite{behera2019magnetic} and Janus magnets \cite{yuan2020intrinsic}, all attributed to the DM interaction, motivating us to study the moir\'{e} physics of such vdW systems.

We will focus on a nontwisted heterobilayer of ferromagnetic (harboring a DM interaction) and antiferromagnetic vdW materials. A similar construction was explored in \cite{Dipolar}, where dipolar interactions along with an external field were added to the moir\'e structure in contrast to the DM interaction of the present work.  {We show that a wealth of different phases featuring magnetic textures on the moir\'e scale and in particular a skyrmion lattice phase could form}.

We will use the continuum formalism introduced in \cite{hejazi2020noncollinear}, which {obviates} the need to consider numerous lattice sites with
complicated local environments. 
First, we review this continuum formalism and derive the Hamiltonian that we will focus on, followed by a perturbative analysis.
The competition between the DM interaction and the moir\'{e} potential induces a commensurate-incommensurate transition that will further be analyzed in the weak-coupling regime. Then 
different types of commensurate phases are introduced, supplemented furthermore by a numerical ground-state phase diagram, 
obtained from a Landau-Lifshitz-Gilbert analysis. Finally, we discuss the possible extensions and experimental relevance of the work.
Details of the weak-coupling analysis will be presented in \cite{supplemental}.

\textit{Setup of the problem}--We study a heterobilayer system of honeycomb ferromagnetic and honeycomb antiferromagnetic vdW materials. This setting is similar to that in \cite{Dipolar}. Both layers are assumed to exhibit long-range order, so a local description in terms of the order parameters $\M$, $\N$ for the ferromagnetic and the N\'{e}el antiferromagnetic layers will be employed, where $\left| \M \right| = \left| \N \right| = 1$. After \cite{hejazi2020noncollinear}, we will develop a continuum model to provide a low energy description of the system. When the interlayer coupling and the displacement gradients act as small perturbations on the intrinsic magnetism of the two layers,
a continuum treatment is justified. 
Since we will only be concerned with ground state configurations (and possible other nearby states) in this work, it suffices to analyze the classical Hamiltonians only. 
This consists of intralayer and interlayer terms; we take the former as 
\begin{equation}
\begin{aligned}
    \mathcal{H}_{\text{intra}} &= \frac{\rho_1}{2} (\nabla \M)^2+\frac{\rho_2}{2} (\nabla \N)^2\\
    &+D \, \M\cdot (\nabla\times \M) - C \, (N_z)^2,
\end{aligned}
\end{equation}
where a single-ion anisotropy for the antiferromagnetic layer and a DM interaction for the ferromagnetic layer are assumed; the latter is predicted to be present in families of Janus magnets in particular \cite{yuan2020intrinsic}. We have utilized a Bloch-type DM interaction in this work, however the same results hold for a Ne\'el-type DM interaction as well \cite{supplemental}. 
The anisotropy term is neglected in the ferromagnetic layer.

We now turn to the interlayer coupling: the spin densities of the two layers are parametrized as:
$\bm{\mathcal{S}}_1= m_0 \bm{M}\left[f_0-\sum_a \cos(\bm{b}_a\cdot \bm{r})\right]$ and $\bm{\mathcal{S}}_2= n_0\bm{N}\sum_a\sin(\bm{c}_a\cdot \bm{r}),$ 
where $\bm{b}_a,\bm{c}_a$ denote the reciprocal lattice vectors of the microscopic lattices in the two layers. Furthermore, $m_0$ and $n_0$ are proportional to the ordered moments in the two layers, and the zeroth Fourier component $f_0$ is undetermined; its value will not matter for the low-energy physics. Taking a minimal Fourier expansion, we limit the summation over $\bm{c}_a$ and $\bm{b}_a$ to the three smallest reciprocal lattice vectors.
Assuming a local interlayer exchange of the form $H'=-J' \bm{\mathcal{S}}_1\cdot \bm{\mathcal{S}}_2$ and that $|\bm{b}_a|-|\bm{c}_b|\ll |\bm{b}_a|,~|\bm{c}_b|$, one arrives at the following interlayer interaction:
$
\mathcal{H}' = J' \M\cdot\N \sum_a \sin(\bm{d}_a \cdot \bm{r}),$
where the three vectors $\bm{d}_a=\bm{b}_a-\bm{c}_a$ represent the reciprocal vectors of an emergent moir\'e lattice, and the fast-oscillating terms have been omitted. We would also like to mention in passing that considering a triangular lattice ferromagnet also leads to the same interlayer coupling.

{$\mathcal{H}=\mathcal{H}_{\text{intra}}+\mathcal{H}'$} describes the system in the general setting described above. However, for simplicity, in the remainder of this work we make the assumption that spin stiffness and anisotropy parameters of the antiferromagnetic layer are large enough to render it essentially nondynamical; 
this results in a uniform N\'eel vector in the $\pm\hat{\bm{z}}$ directions; 
we choose $\N = +\hat{\bm{z}}$. With these assumptions, the Hamiltonian reduces to an effective single-layer one, which, up to a constant rescaling, has the form:
\begin{equation}
\mathcal{H}= \frac12 (\nabla \M)^2+ \beta \, \M\cdot \left(\nabla\times\M \right)+\alpha\, M_z \, \Phi(\bm{x}).
\label{eq:DM_Ham}
\end{equation}
We have introduced the dimensionless parameters
$\alpha=\frac{J'}{\rho_1 d^2},$  $\beta=\frac{D}{\rho_1 d}$
with $d=|\bm{d}_a|$,
and adopted the dimensionless coordinates $\bm{x}=d~\bm{r}$ and defined $\Phi(\bm{x})=\sum_\alpha\sin(\hat{\bm{d}}_\alpha\cdot\bm{x})$.

When $\alpha=0$, ground states of the system consist of spirals with wavenumber $\beta$ (period $2\pi/\beta$). Generically, $\M= \left(\hat{\bm{q}} \times \hat{\bm{z}} \, \sin \left[ \bm{q}\cdot \bm{x} \right], \cos\left[\bm{q}\cdot \bm{x}\right]\right)$ describes a spiral propagating in the $\hat{\bm{q}}$-direction for $\beta>0$ (and $|q|=\beta$). 
In the strong-coupling limit when $\alpha$ is large and $\beta$ is zero, the energy is minimized by a coplanar solution, 
where domains with $\M$ close to $(0,0,\pm 1)$ form, separated by narrow domain walls. Each domain is identified as a region in which $\Phi(\bf{x})$ has a definite sign, resulting in the opposite sign for $M_z$. In Fig.~\ref{fig:wide}a, we plot the structure of $\Phi(\bf{x})$, which forms a triangular lattice; $\Phi(\bf{x})$ has a positive or negative sign in each of the faces of the triangles.

\begin{figure*}[htbp]
 	\centering
    \subfigure[]{\includegraphics[width=0.28\textwidth]{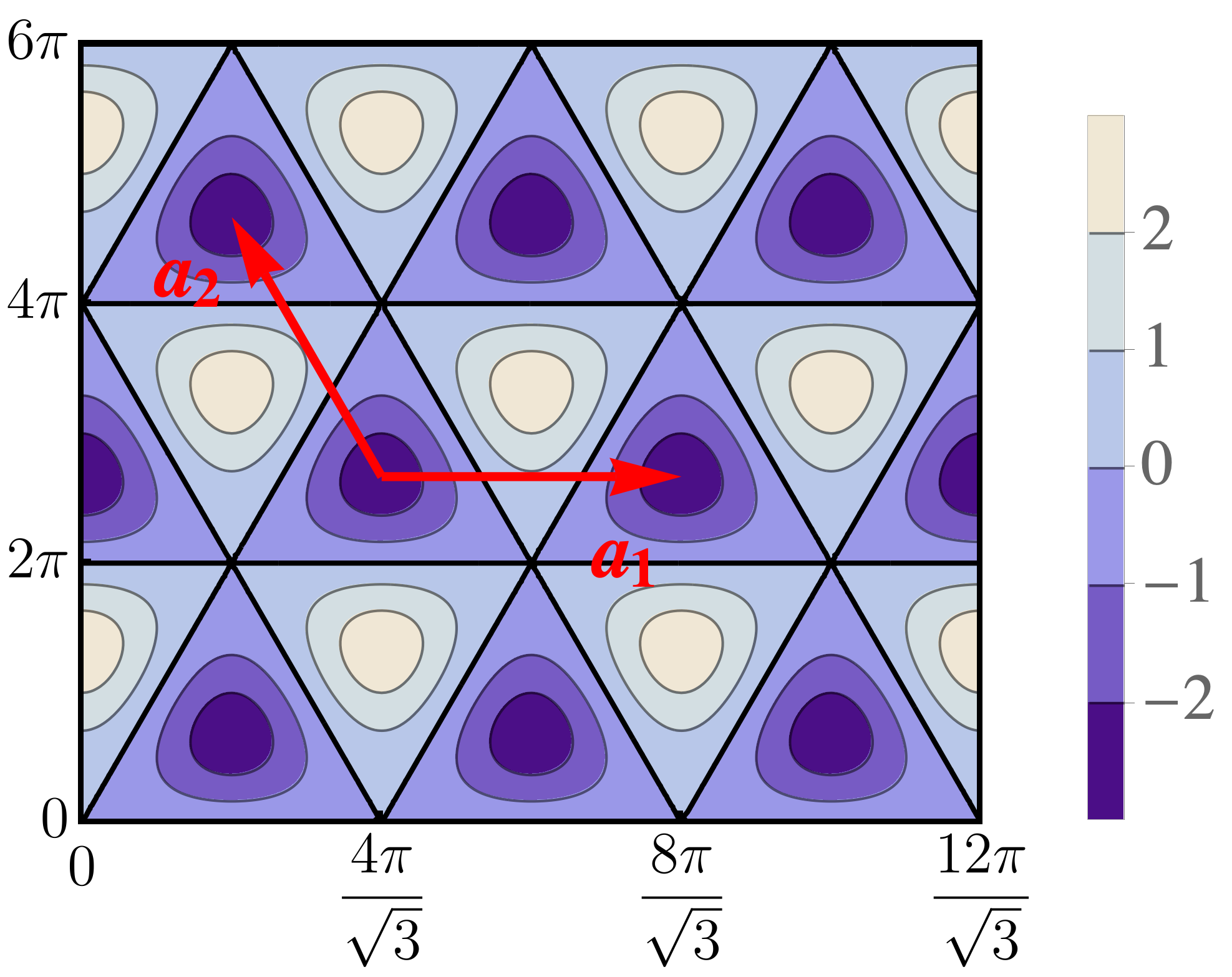}} \quad \
    \subfigure[]{\includegraphics[width=0.33\textwidth]{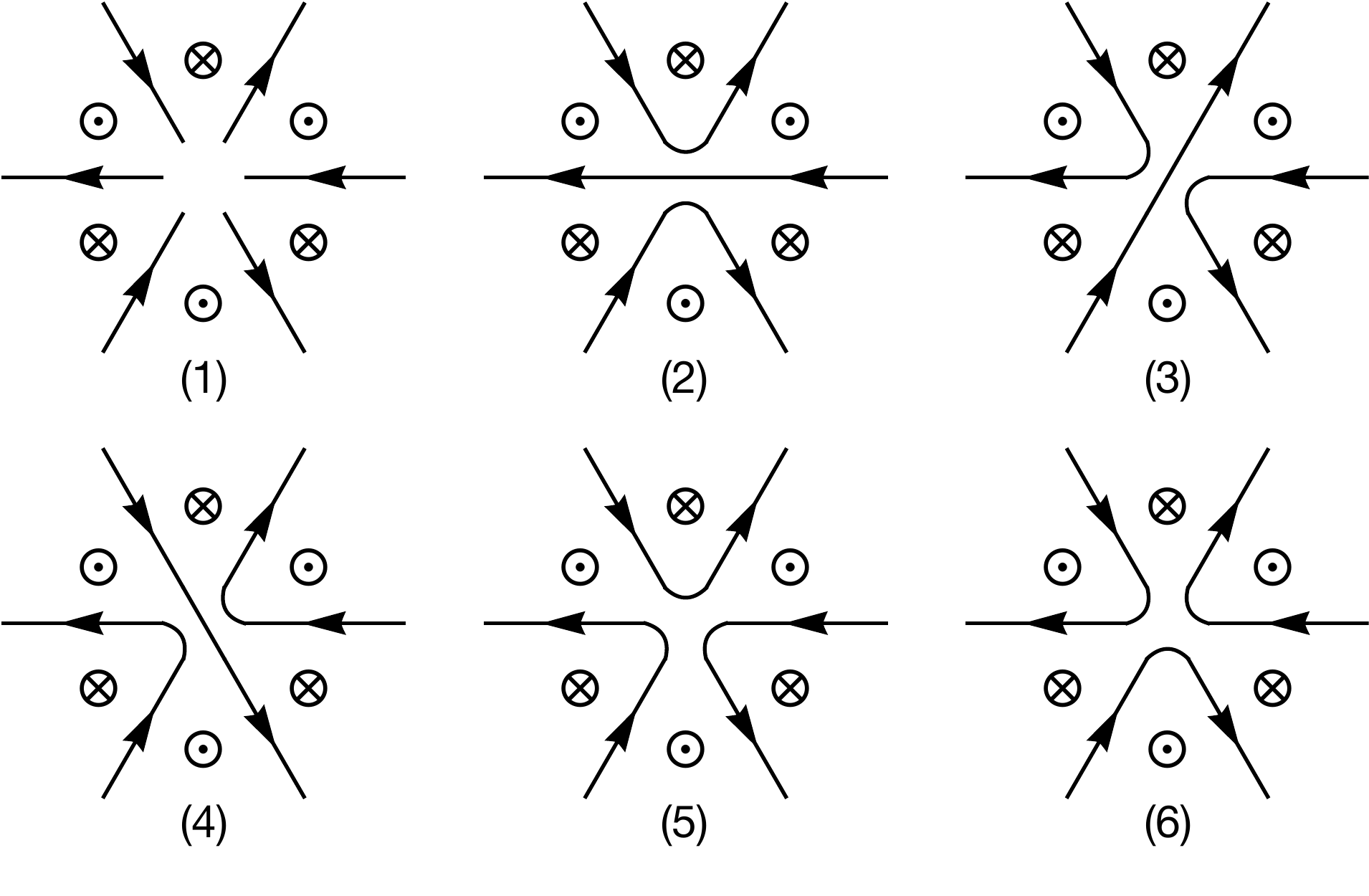}} \quad \
    \subfigure[]{\includegraphics[width=0.33\textwidth]{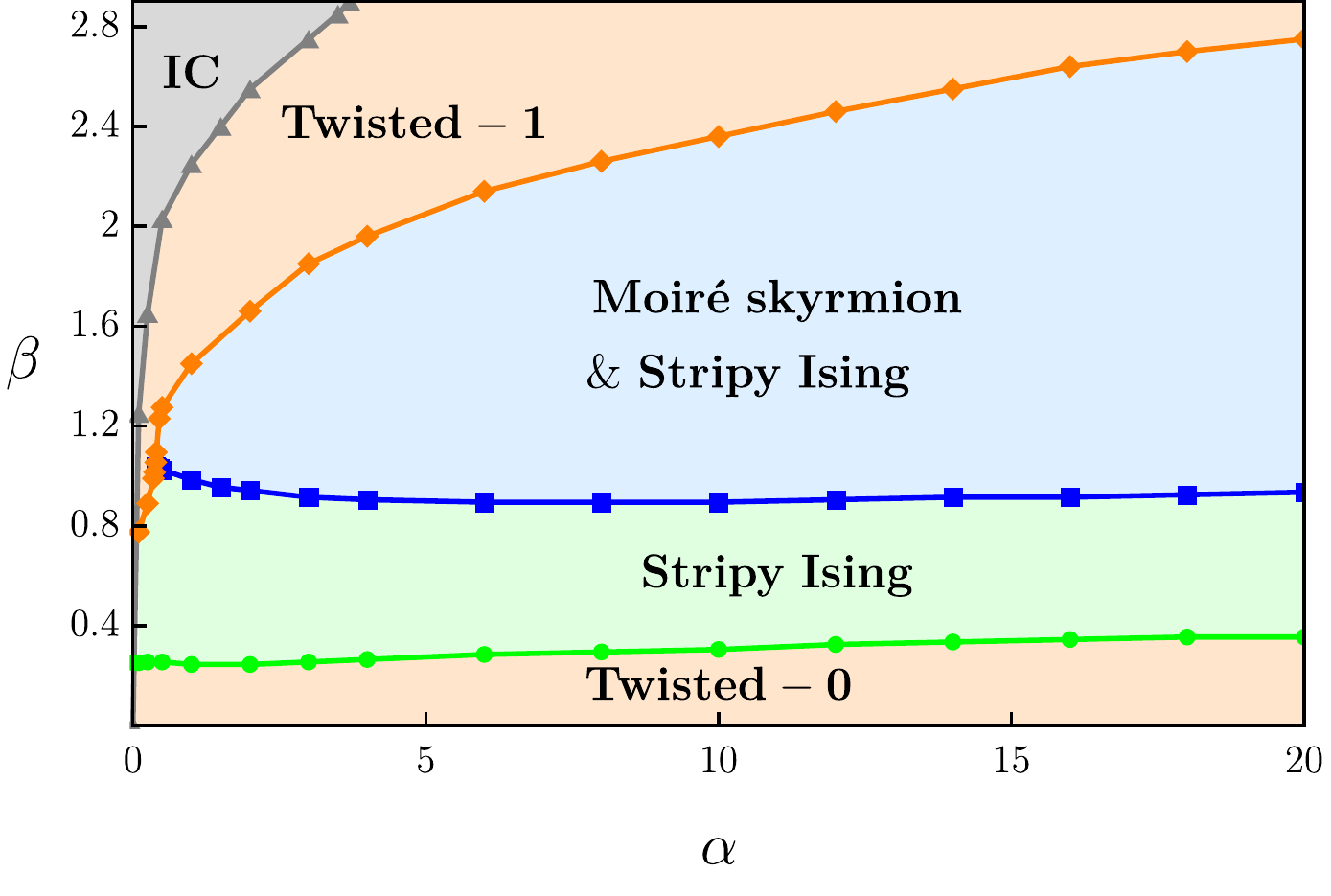}} 
    \caption{(a) The moir\'{e} potential $\Phi(\bm{x})$. $\bm{a}_1$ and  $\bm{a}_2$ form a basis for the emergent moir\'e lattice. 
    (b) (1) An avoided vertex at large $\alpha$ and $\beta$. 
 	The magnetization vectors on the domain walls point along them. 
 	 (2-4) The 3-state clock vertex. 
 	 (5-6) The Ising vertex.
 	 (c) The ground-state phase diagram obtained from LLG equations. 
 	 IC stands for incommensurate phases which can include states of various periodicities. In the blue region of intermediate $\beta$, the moir\'e skyrmion and the stripy Ising states are degenerate in energy. Different commensurate phases are separated by first-order transitions.
 	 \label{fig:wide}
 	 }
\end{figure*}

\textit{Weak-coupling Analysis}--
In the small $\alpha$ and $\beta$ regime, two possibilities can arise: when the $\alpha$ term is dominant over $\beta$, a {\it twisted} solution forms, where in each face of the triangular lattice the magnetization tends to bend towards $-\text{sign}[\Phi(\bm{x})]\hat{\bm{z}}$; nonzero $\beta$ causes distortions in the $\beta=0$ coplanar solution. We call the phase containing these solutions as twisted-0 for a reason that will be explained later.  Alternatively when the $\beta$ term dominates, one expects a spiral configuration perturbed by the moir\'e potential.
Since the spiral can have a period that is different from that of the moir\'e lattice, we call this the incommensurate phase.  A commensurate-incommensurate (CI) transition can occur between these two phases. Below we perform a perturbative solution in the commensurate case, and then extend the analysis to the incommensurate phase and also the transition between the two phases.

We will use a polar coordinates parametrization $\M=(\sin\theta \cos\phi,\sin\theta\sin\phi,\cos\theta)$ for the minimization of the classical Hamiltonian \eqref{eq:DM_Ham}.
 
To zeroth order in $\beta$, the solution to the Euler-Lagrange equation (see \cite{supplemental} for a complete discussion) yields a constant value for $\phi$: $\phi(\bm{x}) = \phi^{(0)} + O(\beta \alpha)$.
The simplified Euler-Lagrange equation for $\theta$ then reads
$\nabla^2\theta + \alpha \sin\theta \, \Phi(\bm{x}) = 0$, with the solution to zeroth order in $\beta$ taking the form $\theta(\bm{x})= \frac{\pi}{2} + \alpha \, \Phi(\x) + O(\alpha^3,\beta^2\alpha)$;
this solution requires a definite mean value for $\theta$, i.e.~$\bar{\theta}=\pi/2$. It furthermore predicts that in the twisted-0 phase, fluctuations of $\theta$ from $\pi/2$ are captured by $\Phi(\bm{x})$ and hence triangular domains form in which spins prefer either up or down directions. We have checked numerically that the solution even with infinitesimal $\beta$, prefers only three values for $\phi^{(0)}$, i.e.~$e^{i\phi^{(0)}} =  \left[\hat{d}_y-i\hat{d}_x \right]^n$, with $n=1,2,3$.

We now turn to a perturbative study of the incommensurate phase and the CI transition; we will focus on the incommensurate solutions close to the CI transition line. 
If $\alpha$ plays a subdominant role, the solution resembles a spiral with a period close to $\frac{2\pi}{\beta}$; this period increases as $\beta$ is lowered towards the CI transition.
Close to the transition point, locally the configuration looks similar to a commensurate one but small-width discommensurations (or solitons) can form with large separations. It is this separation distance that diverges at the CI transition \cite{chaikin1995principles}. We exploit this key information to find incommensurate solutions close to the transition by considering configurations that are locally very close to being commensurate but their local properties could change appreciably if a distance close to the long incommensurate periodicity is taken.

To this end, we make use of the commensurate solution found above, {in which $\theta$ has the mean value $\bar{\theta} = \pi/2$; we now allow this mean value to fluctuate slowly as the position is varied.}  {As a result, we work with a modified Euler-Lagrange equation} $\nabla^2\theta + \alpha \sin\theta \, \Phi(x) = \lambda$, where $\lambda$ is a Lagrange multiplier.
This equation can be solved, and one finds the energy density per unit cell of such a periodic solution to have the form $-\frac34 \alpha^2 \sin^2\bar\theta$; note that this is minimized for $\bar\theta = \pi/2$. Letting $\bar\theta$ fluctuate over long distances will result in an energy penalty; we require this energy penalty to be compensated by the DM interaction energy gain in the incommensurate solutions. In fact an effective one-dimensional Hamiltonian of the form 
$\mathcal{H}^{\text{eff}} = \frac12 \left(\partial_y\bar\theta\right)^2 - \frac34 \alpha^2 \sin^2\bar\theta - \beta \, \partial_y \bar\theta,$
can be exploited to study this competition. This sine-Gordon Hamiltonian has been extensively studied, see e.g.~\cite{chaikin1995principles}, and is known to have a continuous CI transition at $\beta_c=\frac{\sqrt{6}}{\pi} \alpha$.    This means that the phase separation line between the commensurate and the incommensurate phases has a linear form for small $\alpha$ and $\beta$.  For $\beta \gtrsim\beta_c$, the ordering wavevectors deviate from the moir\'e ones (including the vanishing wavevector) with the asymptotic behavior $\delta k \sim  \frac{\sqrt{3} \alpha}{2} \; 1/\log\left( \alpha^2 \frac{1}{\beta - \beta_c} \right)$;
the direction of this incommensurate wavevector correction is not fixed at this order of perturbation theory.

\textit{Possible Commensurate Phases}--
We now turn to the large-$\alpha$ regime where the properties of different commensurate phases are in most contrast. 
In this regime, triangular domains of spins mostly pointing in the $\pm\hat{\bm{z}}$ directions form. 
These domains correspond to those in Fig.~\ref{fig:wide}(a) with sign$[\Phi(x)]=\mp 1$, respectively
The widths of the domain walls (in dimensionless units) decrease as $\alpha$ increases, resulting from the competition of the kinetic and moir\'e potential terms. On the domain walls $\Phi(x)\approx 0$, thus the magnetization is not constrained by the $\alpha$ term. It is actually the different configurations that the magnetization could take on these domain walls that distinguish the
different commensurate phases.
Starting at $\beta=0$, the energy is minimized by a coplanar configuration wherein for all points, the in-plane magnetization can take any direction, i.e.~there is an $SO(2)$ rotational symmetry.

Next we consider the effect of nonzero $\beta$. The DM term, up to integration by parts, can be written as $-2\beta \, \M\cdot (\bm{\hat{z}}\times \bm{\nabla} M_z)$. 
Namely, $\M$ prefers to be perpendicular to the gradient of $M_z$, which lies normal to the domain walls. 
When $\beta$ is not large, it does not affect configurations deep inside the triangular domains, but drives the magnetization vectors on the triangles' edges to point along them in the direction preferred by the DM interaction, as shown in Fig.~\ref{fig:wide}b(1).

However, the configurations close to the vertices are frustrated, i.e., the three directions of magnetization vectors along the edges of the triangles cannot be satisfied at the same time, and hence different vertex configurations could arise. 
One possibility is that the vertex magnetization follows one of the three directions on the edges, breaking the $SO(2)$ symmetry down to $C_3$, as shown in Figs.~\ref{fig:wide}b(2-4); we name this a 3-state clock vertex. 
Interestingly, a solution consisting of the same vertex configuration everywhere has the same symmetries as those of the twisted-0 solution discussed above. We call the phase containing this solution twisted-1 for reasons explained below.

Another possibility of the vertex configuration is Ising-like, i.e.~spins pointing along $+\hat{\bm{z}}$ or $-\hat{\bm{z}}$ as in Figs.~\ref{fig:wide}b(5) and \ref{fig:wide}b(6), enabling the possibility of skyrmions:
if all the vertices choose the same configuration,
say \ref{fig:wide}b(6), a skyrmion lattice of moir\'{e} scale is formed:
triangular domains with up spin constitute a lattice of skyrmions in a down-spin sea (see also Fig.~\ref{fig:different_sols}). 

The skyrmion lattice phase can be viewed as stabilized due to an effective ferromagnetic interaction between neighboring Ising-type vertices, however it could happen that this interaction becomes antiferromagnetic, making a translational-symmetry-broken phase possible: Our numerical results (see below) indicate that one of such phases is possible, where the Ising vertices all point in the same direction along the direction specified by one of the moir\'e lattice unit vectors, while along the directions of other independent unit vectors, the Ising vertices alternatively point up and down (see Fig.~\ref{fig:different_sols}). We name it as the Ising stripy phase. 

{Symmetries of these solutions are discussed in \cite{supplemental}. We go through the full Hamiltonian minimization next and determine the energetically favored solution for different parameter choices.}

\textit{Numerical minimization}--To minimize the energy functional in \eqref{eq:DM_Ham}, we work with the Landau-Lifshitz-Gilbert (LLG) equations \cite{LL,gilbert2004phenomenological}. The variation of the Hamiltonian with respect to the magnetization vector is treated as an effective external field, $\bm{B}^{\text{eff}}=-\delta H/\delta \bm{M}$, which results in the LLG equations: $\frac{d\M}{dt}=-g \, \bm{M} \times \bm{B}^{\text{eff}} + \eta \, \M\times \frac{d\M}{dt}$.
Here $g$ is the gyromagnetic ratio, which is not important since we are only interested in the late-time static configurations, and $\eta$ is the Gilbert damping coefficient. Releasing from various trial configurations and comparing the energies of the final configurations, one can find the ground state with of the Hamiltonian. Fig.~\ref{fig:wide}c shows the numerical phase diagram obtained from LLG equations using the \textit{ubermag} and \textit{OOMMF} packages \cite{ubermag1,ubermag2,oommf}. Different periodic boundary conditions compatible with moir\'e lattice periodicities are imposed. More details on the numerics are presented in \cite{supplemental}.

We find that apart from the incommensurate phases, the twisted-0, the twisted-1, the moir\'e skyrmion and the Ising stripy phases introduced above
constitute the main body of the phase diagram in the strong coupling regime.
In Fig.~\ref{fig:different_sols}, we present examples of steady-state configurations for these three phases (note that the shown configurations are not necessarily ground states for the chosen parameters). In the moir\'{e} skyrmion phase, interestingly, there is one skyrmion per moir\'e unit cell with skyrmion number $\pm 1$ and helicity $\pi/2$, i.e.~$N_{sk}=\frac{1}{4\pi}\iint_{u.c.}d^2\bm{x}~\M\cdot \left(\frac{\partial \M}{\partial x}\times \frac{\partial \M}{\partial y}\right)= \pm 1$; the sign is different for the two types of skyrmion lattices, i.e.~it depends on the direction of magnetization ($+\hat{z}$ or $-\hat{z}$) at the centers of the skyrmions in each solution.
\begin{figure}[htbp]
\includegraphics[width=0.32\linewidth]{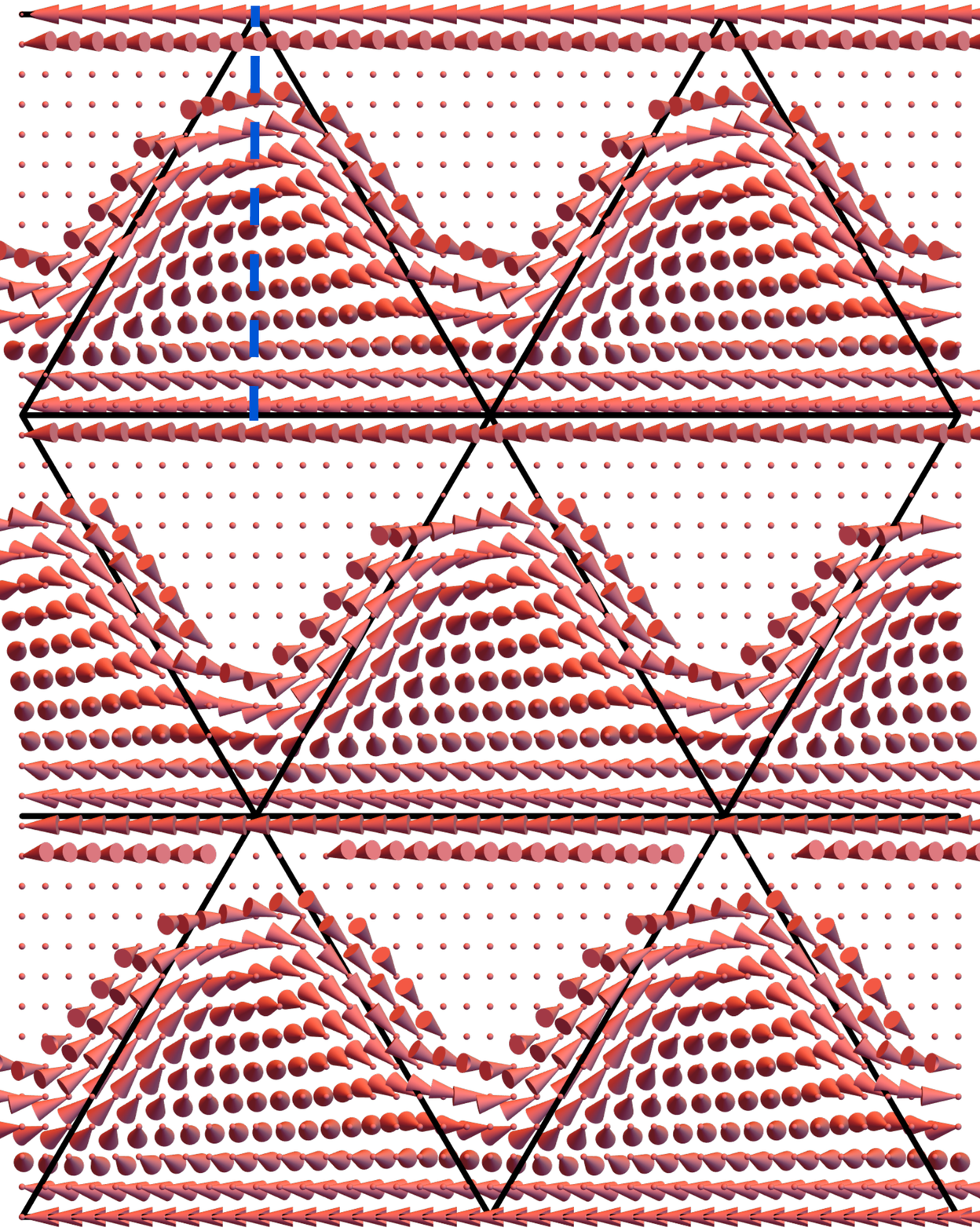}
\includegraphics[width=0.32\linewidth]{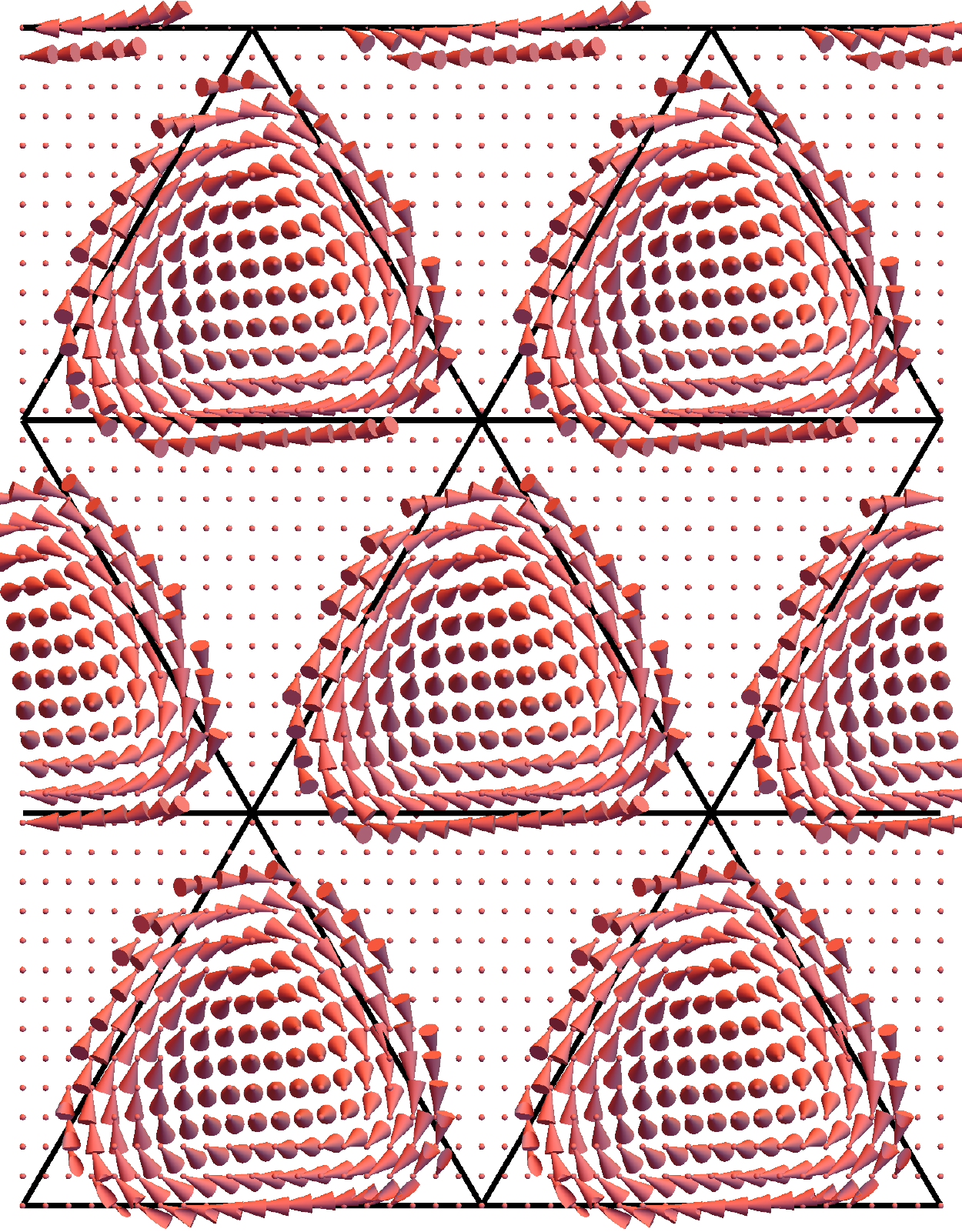}
\includegraphics[width=0.32\linewidth]{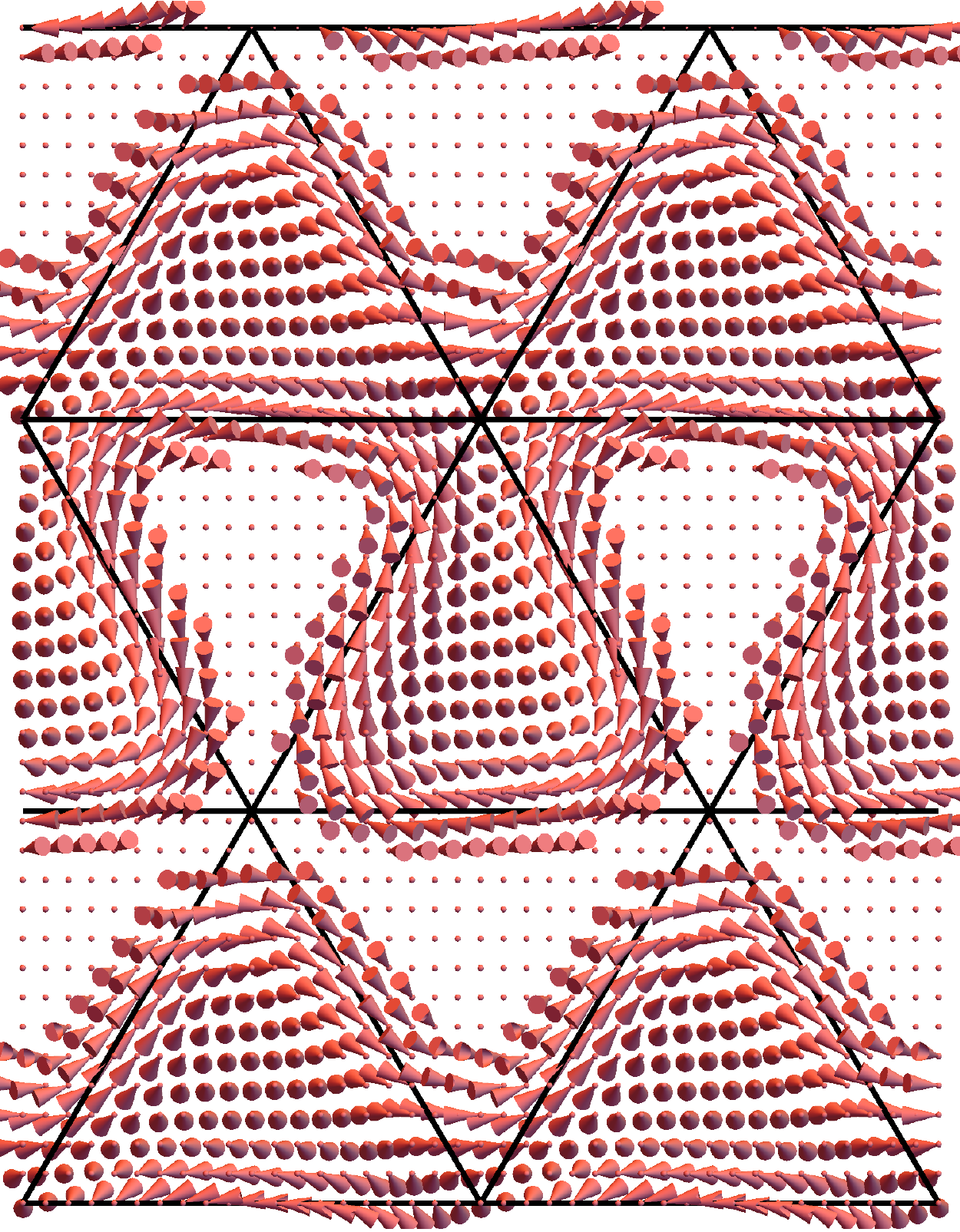}
\caption{Top view of the twisted-1 (left), skyrmion lattice (middle) and the stripy Ising (right) configurations generated at $(\alpha,\beta)=(1.25,1)$, $(1.25,1)$ and $(2.5,0.5)$, respectively. 
The background $-\hat{\bm{z}}$ pointing magnetization vectors are depicted as points. The vertices are of the type shown in Fig.~\ref{fig:wide}b(2) for the twisted-1, of the Fig.~\ref{fig:wide}b(6) type for the skyrmion lattice and of both \ref{fig:wide}b(5) and \ref{fig:wide}b(6) types for the Ising stripy. The solution on the left panel shows a nonzero winding in its magnetization as the dashed line is traversed (see the main text). }
\label{fig:different_sols}
\end{figure}

We see that the twisted-0 phase has the lowest energy for small $\beta$ as expected. Upon increasing $\beta$, it is then replaced by the stripy Ising phase. Interestingly, at larger $\beta$, the stripy Ising and the moir\'e skyrmion states, despite having different symmetries, become degenerate in energy. {Note that these two states both have Ising-type vertices that are only spatially arranged differently.} Then at higher values of $\beta$, the twisted-1 phase, that has vertex configuration \ref{fig:wide}b(2-4) everywhere, minimizes the energy. This phase has the same symmetries as the twisted-0 phase, but is different in the winding numbers along high symmetry lines: 
The magnetization vector rotates exactly by $2\pi$ in the $x-z$ plane along the dashed blue line segment specified in the left panel of Fig.~\ref{fig:different_sols}. On the other hand, the net winding is zero in the twisted-0 phase (see \cite{supplemental} for a strong coupling figure of the twisted-0 phase), hence the names twisted-0 and twisted-1. In principle, other winding numbers can also be be possible (similarly all other solutions can also have higher winding numbers as well). All transitions between these commensurate phases are first order, which can be observed by comparing the energies. More details can be found in \cite{supplemental}. 

In the upper left corner of the phase diagram, we have found different incommensurate phases, i.e., states incommensurate with the moir\'e potential. For example when $\alpha=0$, the periodicity of the spirals is given by $2\pi/\beta$. In general, the competition between $\alpha$ and $\beta$ can lead to configurations with various periods and symmetries, and even chaotic structures \cite{bak1982commensurate}. Since, in the LLG equations, we cannot exhaust all possible trial configurations and unit cells enlargements, and thus the extent of the incommensurate phases is underestimated; as a result, we take a cautious view and do not make definitive conclusions regarding the nature of this region or its phase boundary, except what can be extracted from the weak-coupling analysis presented above. In this weak-coupling regime, the prediction for the CI transition line is supported using numerics with large sample sizes, see the supplemental material \cite{supplemental}.

\textit{Discussion}--A considerable portion of the phase diagram Fig.~\ref{fig:wide}c for $\alpha > 3$ is occupied by phases with the Ising type vertices.
In this paper, we have ignored the single-ion anisotropy for the ferromagnetic layer; such anisotropy, if Ising-like, further stabilizes Ising-type vertices. 
On the other hand, a perpendicular magnetic field will similarly stabilize the Ising vertices, and in particular favor the skyrmion lattice phase over the stripy Ising phase in the blue region of the phase diagram in Fig.~\ref{fig:wide}c; this is consistent with the well-known fact that external fields can stabilize skyrmion lattice phases \cite{bogdanov1989thermodynamically,bogdanov1994thermodynamically,do2009skyrmions, han2010skyrmion, li2011general}.

We have made the assumption that the spin-stiffness and anisotropy parameters of the antiferromagnetic layer are large, such that the antiferromagnetic layer is in the collinear phase. We expect this assumption to be valid in the regime where $C/(\rho_2 d^2)\gtrsim \alpha^2$ and $\rho_2/\rho_1 \gtrsim 1$, as analyzed in our previous work \cite{hejazi2020noncollinear}.  Relaxing this assumption could lead to ground state solutions requiring nontrivial textures in both layers {resembling those in \cite{hejazi2020noncollinear}}.
Furthermore,  {let us mention that} a small twisting between the two layers  effectively increases the magnitude of the wavevector $d$ introduced above. In the dimensionless parametrization, this corresponds to a modification of the spatial structure of $\hat\Phi(\bm{x})$, and a decrease in $\alpha$ and $\beta$.

The material MnPS$_3$, having an antiferromagnetic Heisenberg Hamiltonian and an Ising-like anisotropy \cite{joy1992magnetism,wildes1998spin,okuda1986magnetic,kurosawa1983neutron}, may be a good candidate for the antiferromagnetic layer. The family of Janus transition metal dichalcogenides \cite{zhang2020electronic} could be a promising candidate for the FM layer; these materials have small lattice mismatches compared with MnPS$_3$ and also exhibit the DM interaction \cite{yuan2020intrinsic}.

Interesting phenomena such as the topological Hall effect could arise due to the effect of moir\'e skyrmions on the conduction electrons through the emergent electromagnetic field (see for example \cite{nagaosa2012emergent,nagaosa2012gauge}). 
We leave the full description of the related physics to future work.

\textbf{Note added}: During the preparation of this draft, we noticed \cite{Akram2020} which has some overlap with our work. Their focus is mainly on the LLG simulation of commensurate phases, however, other strong-coupling commensurate phases, incommensurate phases and in particular the commensurate-incommensurate transition studied here are not captured. In particular, the Ising stripy phase which we found to be degenerate with the skyrmion phase is absent their treatment.

\begin{acknowledgments}
{\bf Acknowledgments:} Z.-X. L. is supported by the Simons Collaboration on Ultra-Quantum Matter, which is a grant from the Simons Foundation (651440). L.B. and K.H. were supported by the DOE, Office of Science, Basic Energy Sciences under Award No. DE-FG02-08ER46524. We are grateful to the referee for helpful suggestions on the numerics.
\end{acknowledgments}

\bibliography{DM_PRL.bib}

\vspace{2cm}

\begin{CJK*}{UTF8}{}
\title{{\Large SUPPLEMENTARY MATERIAL} \\
\bigskip
Heterobilayer moir\'e magnets: moir\'e skyrmions and \\ the commensurate-incommensurate transition
}
\CJKfamily{gbsn}

\author{Kasra Hejazi}
\thanks{These two authors contributed equally.}
\affiliation{Department of Physics, University of California Santa Barbara, Santa Barbara, California, 93106-4030, USA}
\author{Zhu-Xi Luo (罗竹悉)}
\thanks{These two authors contributed equally.}
\affiliation{Kavli Institute for Theoretical Physics, University of California, Santa Barbara, CA 93106-4030, USA}
\author{Leon Balents}
\affiliation{Kavli Institute for Theoretical Physics, University of California, Santa Barbara, CA 93106-4030, USA}
\affiliation{Canadian Institute for Advanced Research, Toronto, Ontario, Canada}

\maketitle
\end{CJK*}

\onecolumngrid
\section{Details on weak-coupling analysis} 
\label{sec:weak}
We describe a complete weak-coupling analysis in this section.

Using a polar parametriztion for the magnetization vector $\M=(\sin\theta \cos\phi,\sin\theta\sin\phi,\cos\theta)$, the Hamiltonian density takes the form:
\begin{equation}
\begin{aligned}
\mathcal{H}=& \frac12\left(\nabla\theta \right)^2 + \frac12 \sin^2\theta \, \left(\nabla\phi\right)^2+\alpha \cos\theta \, \Phi(\bm{x})\\
& \qquad \qquad + 2 \, \beta \, \cos\theta \left[\cos\theta \left(\sin\phi \, \partial_x\theta - \cos\phi \, \partial_y\theta \right)
+\sin\theta \left(\cos\phi \, \partial_x\phi + \sin\phi \, \partial_y\phi\right) \right],
\end{aligned}
\end{equation}
where $\Phi(\bm{x})=\sum_\alpha\sin(\hat{\bm{d}}_\alpha\cdot\bm{x})$, and $\bm{x} = \left(x,y\right)$. The Euler-Lagrange equations read:
\begin{equation}
\begin{aligned}
& \nabla^2\theta-\sin\theta\cos\theta\left(\nabla\phi\right)^2+ \alpha\sin\theta \, \Phi(\bm{x})  + 2 \, \beta\sin^2\theta \left(\cos\phi \, \partial_x\phi+\sin\phi \, \partial_y\phi\right)=0,\\
& \sin\theta~\nabla^2\phi + 2\cos\theta \left(\nabla\phi\cdot\nabla\theta \right) - 2 \, \beta\sin\theta \left(\cos\phi \, \partial_x\theta+\sin\phi \, \partial_y\theta\right)=0.
\end{aligned}
\end{equation}
First, we seek a commensurate solution for $\theta$ and $\phi$; the perturbation expansion to lowest orders in $\beta$ and $\alpha$ read:
\begin{equation}
\label{eq:perturb_solutions_tetha_phi}
\begin{aligned}
    \theta &= \frac{\pi}{2} + \alpha \, \Phi(\bm{x}) + O(\alpha^3) \\
    & \qquad + 4 \beta^2\alpha \left[ \cos^2\phi^{(0)} \, \partial_x^2 \Phi(\bm{x}) + 2\sin\phi^{(0)}\cos\phi^{(0)} \, \partial_x\partial_y \Phi(\bm{x}) + \sin^2\phi^{(0)} \, \partial_y^2 \Phi(\bm{x})  \right] + \ldots, \\
    \phi &= \phi^{(0)} - 2\beta \alpha \left( \cos\phi^{(0)} \partial_x\Phi(\bm{x}) + \sin\phi^{(0)} \partial_y\Phi(\bm{x})  \right)  + \ldots \ .
\end{aligned}
\end{equation}
The energy per unit cell taking the solutions up to this order into account reads:
\begin{equation}\label{eq:energy_density_commensurate}
\begin{aligned}
    \mathcal{E} &= - \frac{3}{4} \alpha^2 + \ldots \\
    &\quad -2 \beta^2 \alpha^2 \sum_a \left( \hat{\bm{m}}_\parallel^{(0)} . \hat{\bm{d}}_a \right)^2 + \ldots,
\end{aligned}
\end{equation}
where $\hat{\bm{m}}_\parallel$ is a unit vector in the direction of the in-plane component of the magnetization vector $\bm{M}$. The $\ldots$ on the first row stands for terms higher order in $\alpha$ and zeroth order in $\beta$, while the $\ldots$ on the second row represents higher orders in both $\alpha$ and $\beta$. The above form can be further simplified using $\sum_a \left( \hat{\bm{m}}_\parallel^{(0)} \cdot \hat{\bm{d}}_a \right)^2 = \frac32$, this means that to this order there is no preferred direction for $\hat{\bm{m}}_\parallel^{(0)}$ or equivalently that $\phi^{(0)}$ is not determined to this order. We expect higher order corrections to break this rotational symmetry and give $\phi^{(0)}$ the three preferred values (that are $C_3$ equivalent) we found numerically. 

Now we turn to a perturbative study of incommensurate configurations close to the commensurate-incommensurate transition line. Such configurations can show slow variations of the magnetization over a large {\it incommensurate length scale} on top of the fluctuations on the moir\'e scale.
Additionally, one can see that the above form for the average energy per unit cell  \eqref{eq:energy_density_commensurate} contains no linear-in-$\beta$ contribution; however such a linear term can arise in an incommensurate configuration where the angle $\theta$ acquires some accumulated winding over a long length. In the following, we will study this contribution and how it can lead to the commensurate-incommensurate transition.

In this limit, the following approximation is made: we take the incommensurate configuration to look locally like a commensurate one for which we found the perturbative solution above. In the above solution, the mean value of $\theta$ in moir\'e unit cells has to take the value $\pi/2$ in order for the energy to be minimized. However, in order to allow for large scale variations in $\theta$, we let its mean value to change slowly as one moves in the moir\'e lattice; this will bring in some energy penalty due to the fact that in some unit cells the mean value of $\theta$ is different from its preferred value. The requirement for this energy penalty to be compensated by the linear-in-$\beta$ term (arising from large-scale variations of $\theta$ also) allows us to find the point where the commensurate-incommensurate transition occurs.

To this end, we modify the above solution to accommodate an arbitrary mean for $\theta$ in each unit cell. This can be done by adding a Lagrange multiplier term to the Hamiltonian that is minimized. To zeroth order in $\beta$, $\phi = \text{const.} + O(\beta)$, and thus an equation of the form $\nabla^2\theta + \alpha \sin\theta \, \hat{\Phi}(x) = \lambda$ is obtained. $\lambda$ is the Lagrange multiplier and is found order by order to ensure that $\bar{\theta}$ is the mean of $\theta$ to all orders. Solving with a definite $\bar{\theta}$:
\begin{equation}
    \theta(\bm{x}) = \bar\theta + \alpha \sin\bar\theta \, \Phi({\bm{x}}).
\end{equation}
The energy per moir\'e unit cell is derived using the Hamiltonian density given by $\frac12 \left(\nabla\theta \right)^2 + \alpha \cos\theta \, \Phi(\bm{x})$:
\begin{equation}
\mathcal{E} = -\frac34 \alpha^2 \sin^2\bar\theta . 
\end{equation}
 $\bar\theta = \pi/2$ minimizes this energy density and thus we expect that in the twisted solution fluctuations of $\theta$ happen around $\bar\theta = \pi/2$ as shown in \eqref{eq:perturb_solutions_tetha_phi}. 

\begin{figure}[!t]
\includegraphics[width=0.5\textwidth]{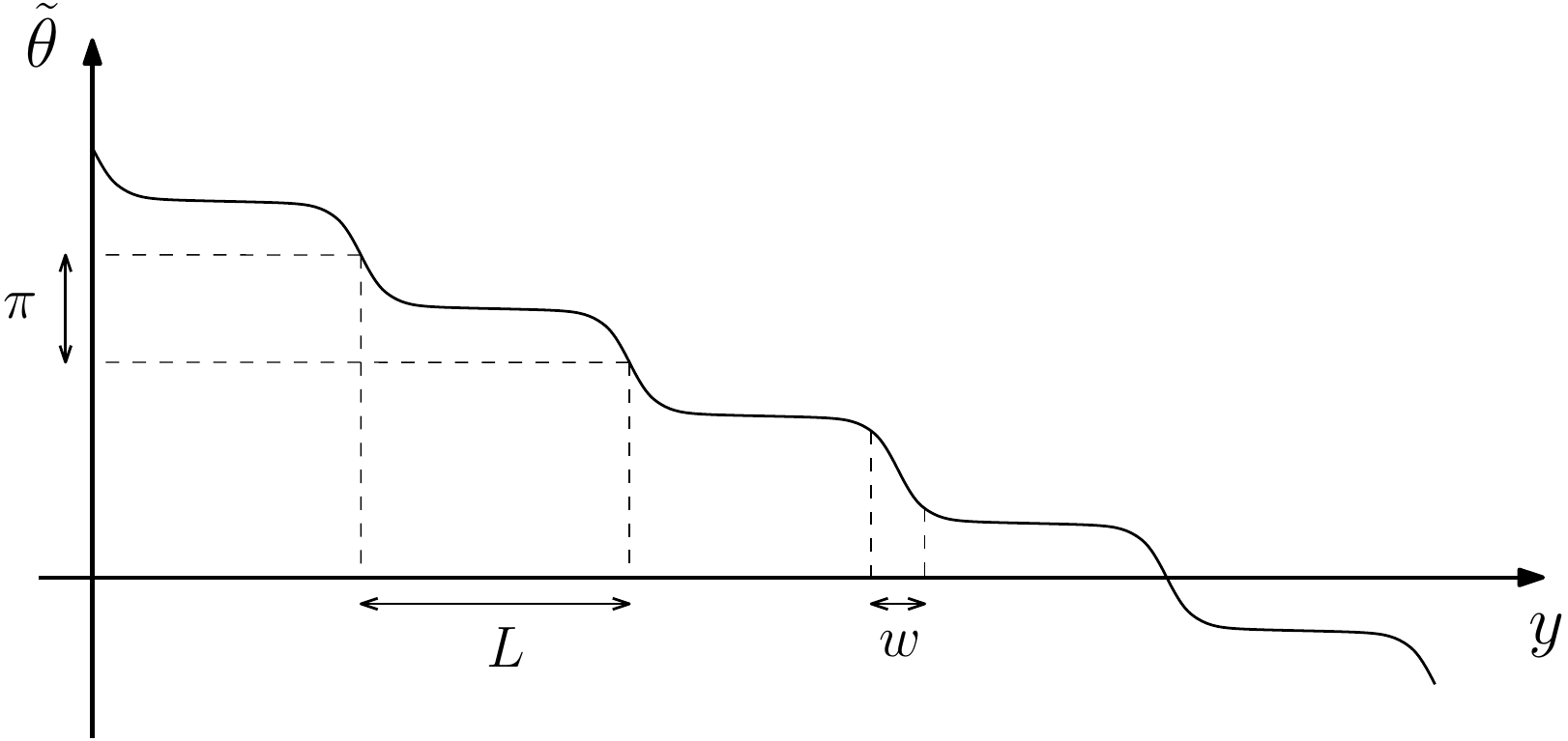}
\caption{\footnotesize A possible configuration for the variable $\tilde\theta = \pi/2 - \bar\theta$ in the incommensurate phase close to the CI transition, where $\bar\theta$ is the local mean value of the $\theta$ angle in unit cells. An incommensurate configuration consists of large regions with $\bar\theta$ close to $\frac{\pi}{2}$ which minimizes the energy in a commensurate configuration, however these regions are separated by narrow discommensurations (or solitons) where $\bar\theta$ varies by $\pi$. The width of the discommensurations is shown by $w$ and their relative distance with $L$. As one gets closer to the transition line $L$ diverges. This plot is inspired by a figure in Ref.~\cite{chaikin1995principles}. }
\label{fig:CI_transition}
\end{figure}

Now one can let $\bar\theta$ to fluctuate slowly from a unit cell to another one that has an appreciable distance, such fluctuations will be controlled by an effective Hamiltonian of the form $\mathcal{H}^{\text{eff}}_{\text{kin+pot}}= \frac12 \left(\nabla\bar\theta\right)^2 - \frac34 \alpha^2 \sin^2\bar\theta$, where the gradient is discretized and $\bar\theta$ takes values in moir\'e unit cells and varies slowly with position. One also needs to add the effect of the DM interaction, to first order in $\beta$, the Hamiltonian density reads $\mathcal{H}^{\text{eff}}_{\text{DM}} = 2\beta \cos^2\bar\theta (\sin\phi \, \partial_x \bar\theta - \cos\phi \, \partial_y \bar\theta)$,  where we have used the fact that $\phi$ is a constant to zeroth order in $\beta$. Any choice of $\phi$ determines a preferred direction for $\bar\theta$ variations. For example $\phi=0$ corresponds to having variations of $\bar\theta$ in the $y$ direction only. To this order in perturbation theory, there is no preferred direction for the incommensurate wavevector. Proceeding with this choice of $\phi$, an effective one-dimensional Hamiltonian, taking all the above points into account, can be achieved:
\begin{equation}
    \mathcal{H}^{\text{eff}} = \frac12 \left(\partial_y\bar\theta\right)^2 - \frac34 \alpha^2 \sin^2\bar\theta - \beta \, \partial_y \bar\theta,
\end{equation}
the Hamiltonian is defined on a dicretized lattice but we use a continuum limit approximation, which is most justified in the simultaneous limits of small $\alpha$ and $\beta$ and large incommensurate periodicity. With a transformation $\tilde{\theta} = \pi/2 - \bar\theta$, the potential term in the effective Hamiltonian takes a positive definite form and one arrives at $\mathcal{H}^{\text{eff}} = \frac12 \left(\partial_y\tilde\theta\right)^2 + \frac34 \alpha^2 \sin^2\tilde\theta + \beta \, \partial_y \tilde\theta$. Such Hamiltonians and their commensurate-incommensurate transitions are studied extensively in Ref.~\cite{chaikin1995principles}. In particular, it is shown that a profile like the one shown in Fig.~\ref{fig:CI_transition} is expected for the variable $\tilde\theta$ in the incommensurate phase; the transition happens at a $\beta$ value equal to $\beta_c = \frac{\sqrt{6}}{\pi} \alpha$, and above this $\beta$ value, creation of the solitons shown in Fig.~\ref{fig:CI_transition} becomes energetically favored; in fact their relative distance is also found to have a form like $L = \frac{2}{\sqrt{3} \alpha} \log\left( \alpha^2 \frac{1}{\beta - \beta_c} \right)$. Notice that in reality, the value of $L$ is bounded by the sample size from above, and the transition will happen at a larger $\beta>\beta_c$.

\begin{figure}[t]
\centering
\includegraphics[scale=0.25]{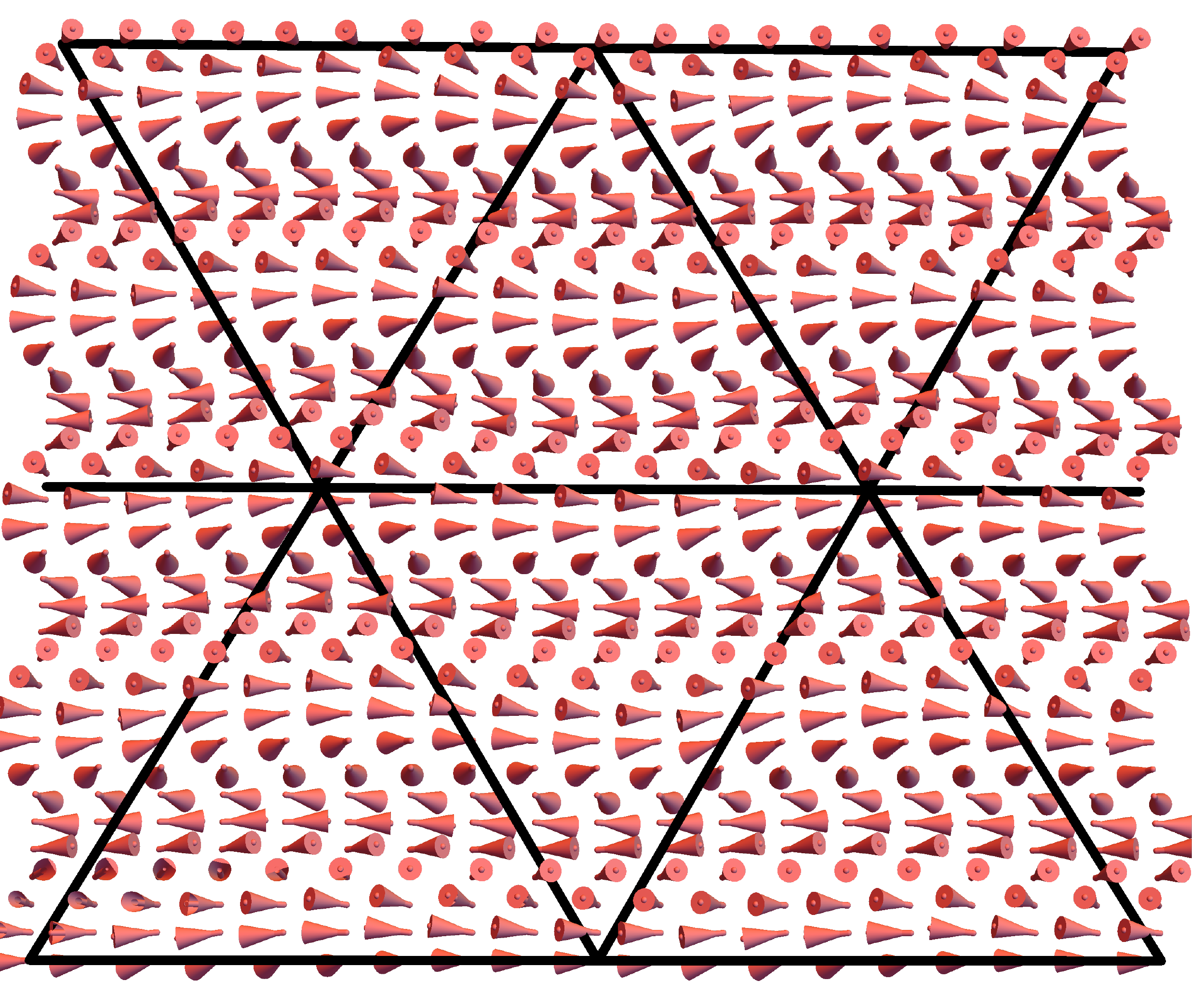}
\caption{The incommensurate relaxed spiral configuration at $\alpha=0.3$ and $\beta=2.2$ with incommensurate periodicity of $4\pi/\beta$ in the $y$-direction. To make explicit the incommensuration, we draw the moir\'e triangular lattices in black lines for comparison.}
\label{fig:incommensurate}
\end{figure}

A plot of an incommensurate solution with a dominant $\beta$ term is presented in Fig.~\ref{fig:incommensurate}, where the numerical simulation is initialized with a spiral configuration assuming $\alpha = 0$, and the effect of the moir\'e potential with $\alpha \neq 0$ is implemented in the numerical minimization of the energy, which will result in intracell modulations of the spiral configuration. The period of the incommenurate solution is taken to be given only in terms of $\beta$, i.e.~having the value $2\pi/\beta$; one expects the period to also have corrections due to the nonzero value of $\alpha$, but since $\alpha$ is chosen to be small, we expect this correction to be perturbative and also expect the general properties of the solution to be unaltered. Such solution is far from the transition line between commensurate and incommensurate solutions, and thus not resembling the perturbative solution of the incommensurate phase as discussed above. However, notice that both of these solutions lie on the incommensurate side of the transition line.

\begin{figure}[htbp]
\centering
\includegraphics[scale=0.5,trim={1cm 0 0 0}, clip]{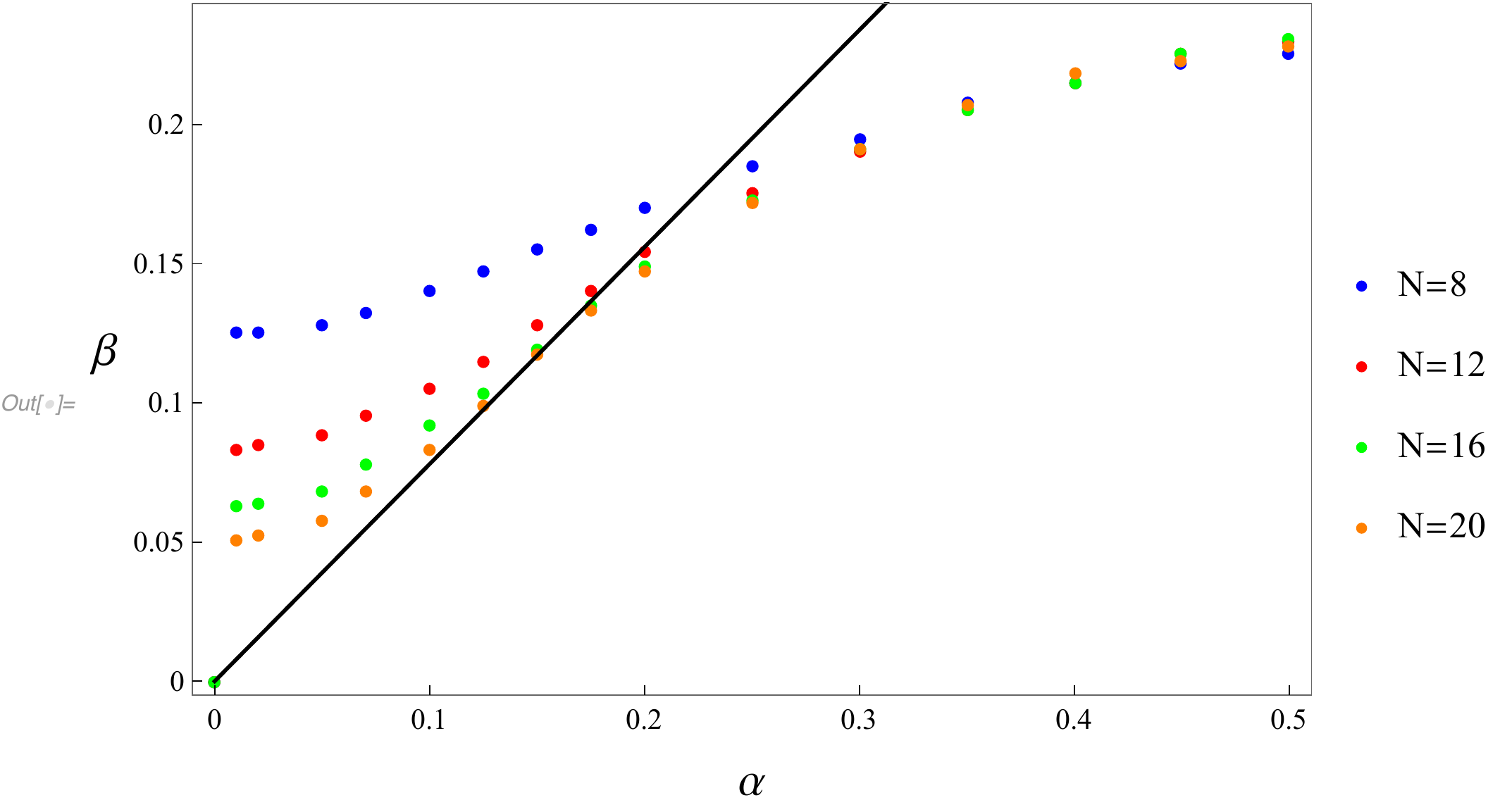}
\caption{Comparison between the theoretically predicted and the numerically found CI transition lines. The numerical results are based on periodic boundary conditions with $N$ moire unit cells. The system is chosen to consist of 2 unit cells in one direction, and as a result $N/2$ unit cells in the other direction. The solution in each case accommodates only a winding equal to $1$ in the larger direction of the system.  As the system size increases from $N=8$ moir\'e unit cells to $20$ moir\'e unit cells, the numerical result approach the theoretical prediction. The numerical lines deviate from theory at larger values of $\alpha$ and $\beta$ because the effects of higher-order terms begin kicking in in that region. The deviations at near the origin result from the fact that at tiny $\beta$, the relaxed spirals have huge periodicity, while the sample size is kept finite.}
\label{fig:match}
\end{figure}

One can approach the transition line by considering relaxed spirals in larger systems, and also letting their period vary. In figure \ref{fig:match}, we provide a comparison between the theoretically predicted CI transition line of $\beta_c = \frac{\sqrt{6}}{\pi} \alpha$ and the numerically found transition lines at various sample sizes. 
As the system size increases, for small enough $\alpha$ (but not too small, see below), the numerical line approaches the theoretical prediction which is based on infinite sample size and found perturbatively for small $\alpha$ and $\beta$. Note that for very small $\alpha$, the numerically found $\beta$ values for the transition show a saturation which is due to the finite size of the numerics. Furthermore, the saturation value decreases as the sample size increases.

\section{Details on numerics}
In this section, we give more details on the Landau-Lifshitz-Gilbert simulation and the numerical phase diagram presented in figure 4. In the LLG equations 
$\frac{d\M}{dt}=-g\M\times \bm{B}^{\text{eff}}+\eta \M\times \frac{d\M}{dt}$, 
can be rewritten as $$\frac{d\M}{dt}=-\frac{g}{1+\eta^2} \M\times \bm{B}-\frac{\eta g}{1+\eta^2}\M\times(\M\times \bm{B}). $$ For faster convergence, we ignore the torque term  $\M\times\bm{B}$ and the simulation evolves towards the equilibrium along a steepest descent path.  
Our boundary conditions include commensurate periodic boundary conditions $\M(x,y)=\M(x+L_x,y)=\M(x,y+L_y)$ with several different choices of $(L_x, L_y)$: such that the sample consists of $2\times 2$, $2\times 1$ and $1\times 1$ moir\'e unit cells 
in the orthogonal basis $\tilde{\bm{a}}_1=(4\pi/\sqrt{3},0),$ $\tilde{\bm{a}}_2=(0,4\pi)$ respectively (or $2\times 4$, $2\times 2$ and $1\times 2$ unit cells in the non-orthogonal basis defined in figure 1).
Since we are mainly interested in the configurations commensurate with the moir\'e potential, a large number of moir\'e unit cells is not necessary. In general, the states with larger sample sizes typically have higher averaged energy densities because of minor breakings of the moir\'e translation symmetries $\M(\x)=\M(\x+\bm{a}_1)=\M(\x+\bm{a}_2)$ allowed by the boundary conditions of larger periods. 

\begin{figure}[htbp]
\centering
\includegraphics[trim=1cm 0 0 0, clip,scale=0.5]{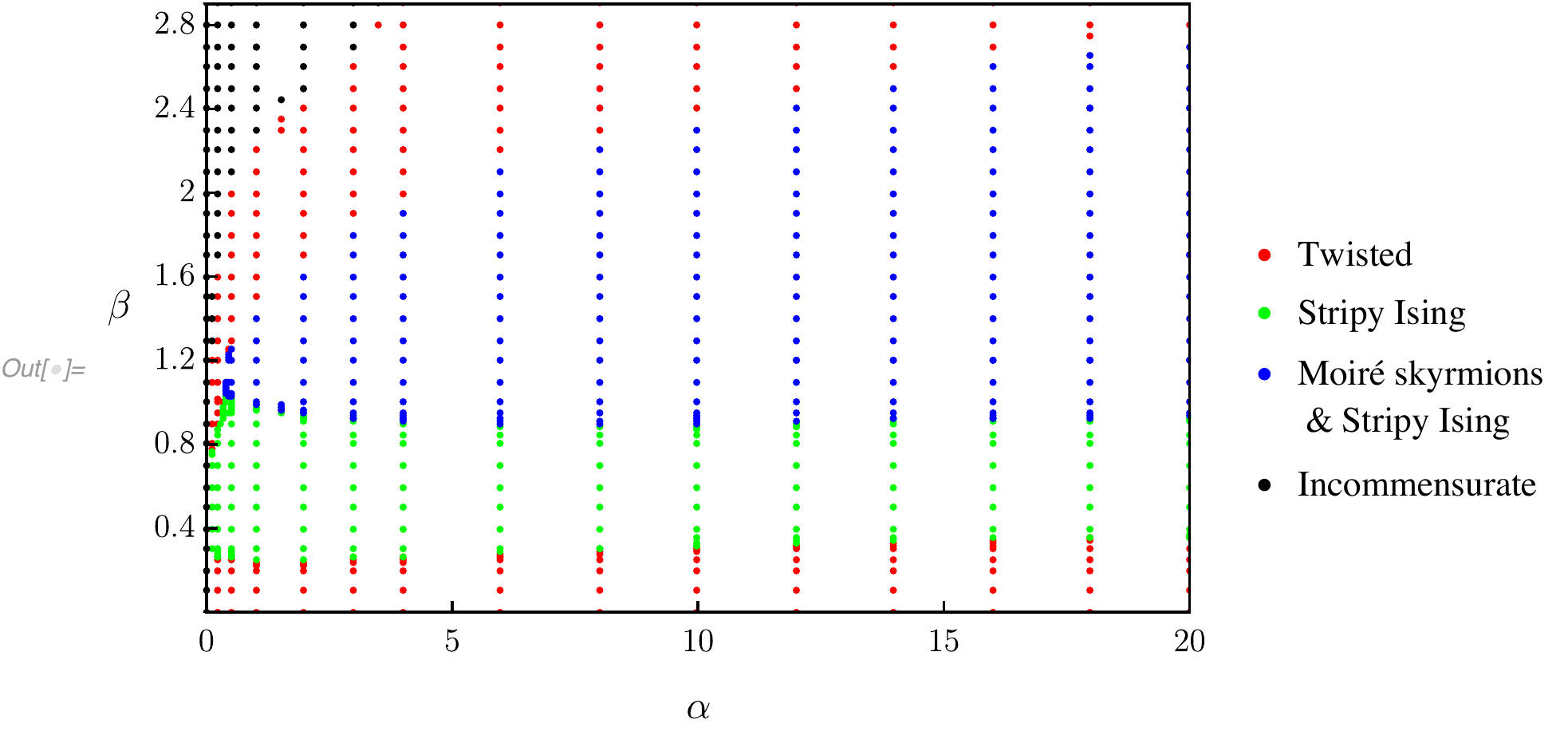} 
\caption{The set of all parameters we choose for the phase diagram.}
\label{fig:points}
\end{figure}

The initial conditions we take are the following:
\begin{itemize}
    \item[(1)] Uniform magnetization in the $\hat{\x}$-direction, $\M(\x)=\hat{\x}$.
    \item[(2)] Uniform magnetization in the $\hat{\bm{z}}$-direction, $\M(\x)=\hat{\bm{z}}$. The uniform configurations are ground states when there is no DM interaction or moir\'e potential, i.e., $\alpha=0=\beta$.
    \item[(3)] Twisted configuration along the $\pm \hat{\bm{z}}$ directions, $\M(\x)=-\text{sign}[\Phi(\x)]$, where $\Phi(\x)$ is the moir\'e potential. This state has the lowest energy when $\beta=0$ and $\alpha$ is finite. 
    \item[(4)] Spiral configuration propagating along the $\hat{\bm{y}}$-direction, $\M(\x)=\sin (\beta y)\hat{\bm{x}}+\cos (\beta y)\hat{\bm{z}},$ which is the ground state when $\alpha=0$ and $\beta>0$. 
    \item[(5)] The skyrmion lattice configuration with the scale characterized by $\beta$. It is generated from summation of three spirals propagating in three different directions.
    \item[(6)] The commensurate twisted state which breaks the $SO(2)$ symmetry down to $C_3$, with the vertex configuration of the type (b)(c)(d) in figure 2. The configuration pattern is the same as that in figure 3 (left).
    \item[(7)] The commensurate stripy Ising state with the vertex configuration (e)(f) in figure 2. The configuration pattern is the same as that in figure 3 (right).
    \item[(8)] The commensurate moir\'e skyrmion lattice with the vertex configuration (e)(f) in figure 2. The configuration pattern is the same as that in figure 3 (middle).
\end{itemize}
The LLG equations are then solved with \textit{OOMMF} using the \textit{ubermag} framework. The maximal step size is $L_i/200$ in the two directions away from the phase boundaries and $L_i/300$ near the phase boundaries. After the final configurations are obtained, we compute the corresponding skyrmion numbers and averaged energy densities and plot the phase diagram. The phase diagram is generated using 668 data points with more points distributed near the phase boundaries. Figure \ref{fig:points} shows the distribution of parameters that we choose. 

Below we show the energy (density) differences between the twisted state and the stripy Ising state, and between the stripy Ising state and the moir\'e skyrmion state, respectively. The phase transitions are thus clearly first-order.
\begin{figure}[htbp]
    \centering
    \includegraphics[scale=0.5]{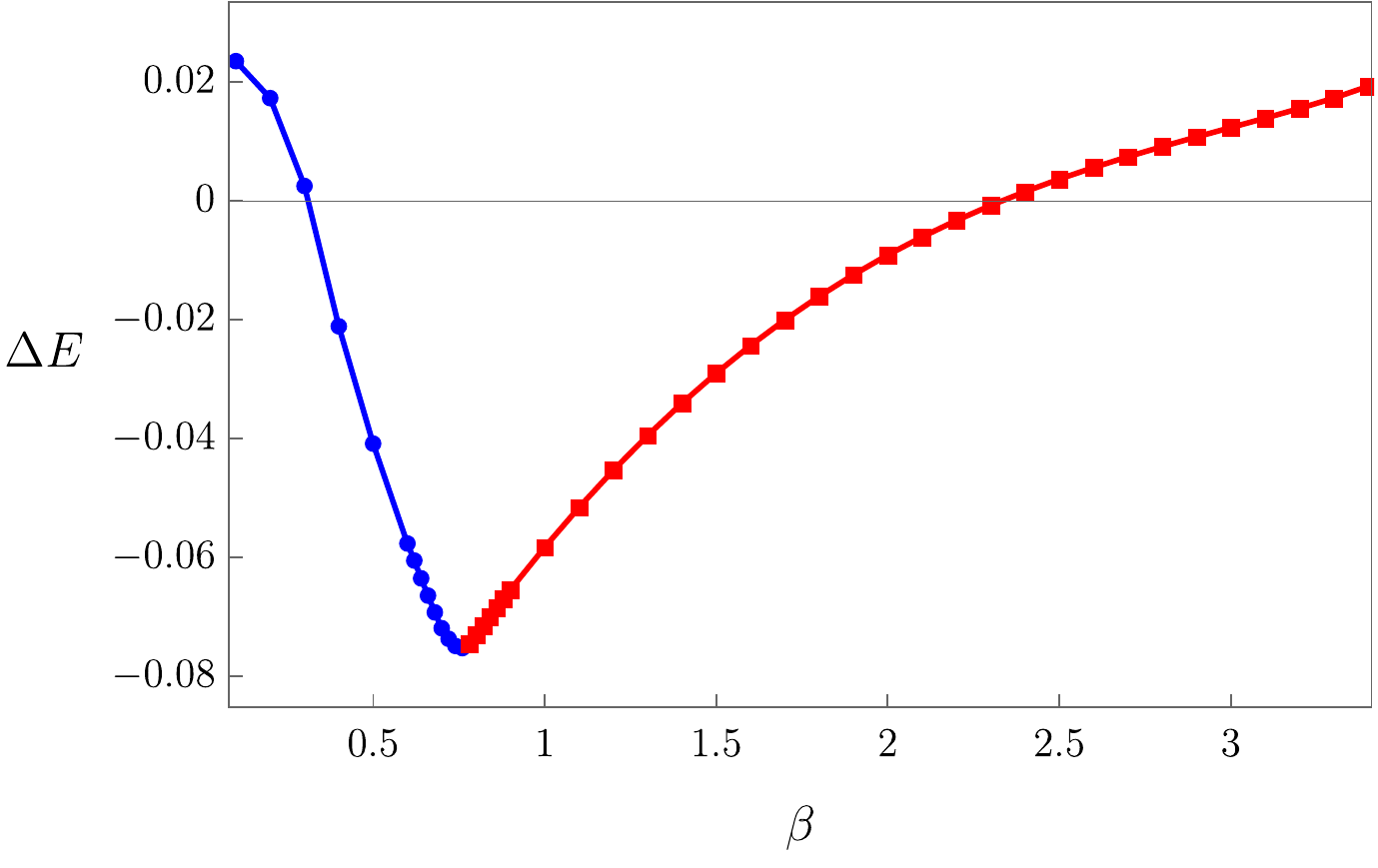}
    \includegraphics[scale=0.5]{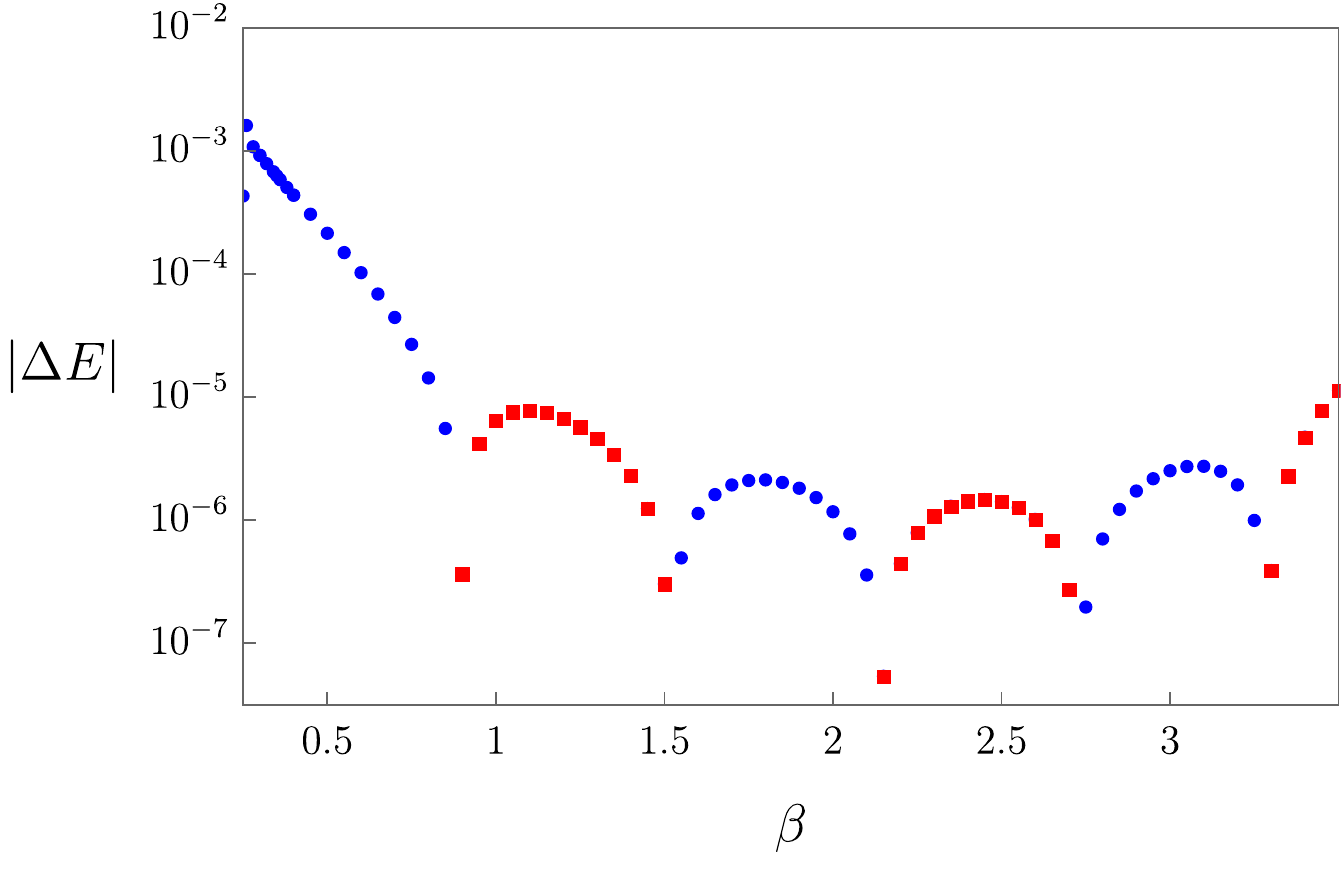} 
    \caption{Energy density differences of the commensurate phases at $\alpha=10$ and various $\beta$. Left: $E_{\text{stripy}}-E_{\text{twisted}}$. The blue dots correspond to the twisted-0 solution at lower $\beta$, while the red dots correspond to the twisted-1 solution at higher $\beta$. 
    Right: $|E_{\text{skyrmion}}-E_{\text{stripy}}|$. Red/blue dots correspond to the cases where $E_{\text{skyrmion}}$ is larger/smaller than $E_{\text{stripy}}$, respectively. We view the two states as degenerate when the magnitude of the energy density difference between the two states is smaller than $10^{-5}$. At smaller $\beta$, the stripy Ising state is preferred, while at larger $\beta$ they become degenerate. }
    \label{fig:energy}
\end{figure}

\begin{figure}[htbp]
    \centering
    \includegraphics[scale=0.25]{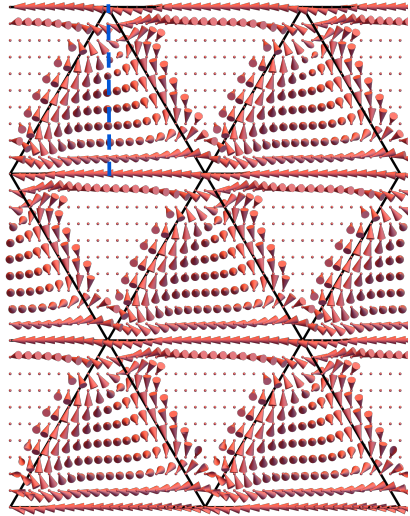}
    \caption{The twisted-0 solution at the strong coupling ($\alpha = 10, \beta =  0.6$), it harbors a trivial configuration at all its vertices. The fact that all the vertices are trivial makes the winding number along the high symmetry line specified with the dashed line vanish. This is in contrast with the twisted-1 phase where the corresponding winding number is 1.}
    \label{fig:twisted_0}
\end{figure}

Finally, we comment on the incommensurate configurations. In the top left region of the phase diagram of large $\beta$ and small $\alpha$, the configuration with the lowest energy (generated from the initial conditions listed above) is that of the relaxed spiral configuration. The rigid spiral configuration was described in the main text above the ``weak-coupling analysis'' section, with magnetization vectors rotating in one plane as the spiral propagates. The relaxed spiral configuration is roughly the same as the rigid spiral configuration, but with small modifications due to the moir\'e potential. The fact that such relaxed spiral configuration has lower energy than the other commensurate configurations, means that the effect of $\beta$ is dominant over that of $\alpha$. Therefore, potentially there could be other configurations with different periodicities that are incommensurate with the moir\'e potential and have even lower energies. This periodicity is expected to change smoothly with $\alpha$ and $\beta$. Since numerically one can only implement boundary conditions with a finite set of different periodicities, the lowest-energy configuration obtained from these boundary conditions is likely not the true ground state of the system.
In particular, the region occupied by the incommensurate phases is underestimated. We thus focus only on the commensurate phases in the main phase diagram that are more predictable. We also perform a numerical analysis in the weak-coupling regime with large sample size and the results show a nice match with the theoretical prediction, see the section ``Details of weak-coupling analysis'' of this supplemental material for details.

\section{Symmetries of the commensurate solutions}
In this section, we discuss the symmetries of the three different commensurate solutions discussed in the main text. The fact that they possess different symmetries means that they belong to different phases. We focus on strong coupling realizations below for clarity:
\begin{itemize}
    \item \textbf{Twisted phases:} The twisted-0 and the twisted-1 phases have the same symmetries. Consider for example a twisted-1 solution  with e.g.~the 3-state clock vertex configuration named as (b) in the main text Fig.~1 (center). Suppose such configuration is realized at all of the vertices, its symmetries include: translation symmetries of the moir\'{e} triangular lattice $T_{\bm{a}_1}$, $T_{\bm{a}_2}$ (the basis vectors are also shown in Fig.~1 (left)), reflection symmetry with respect to the $y=0$ line, i.e. a simultaneous action of $y\rightarrow -y$ and $M_z\rightarrow -M_z$. 
    \item \textbf{Skyrmions phase:} Symmetries of this phase include again the transnational ones $T_{\bm{a}_1}$, $T_{\bm{a}_2}$, and there is an additional $C_3$ rotation symmetry.
    \item \textbf{Stripy Ising phase:} The symmetries are: $T_{\bm{a}_1}$, $T_{2\bm{a}_2}$, and a simultaneous action of $y\rightarrow -y$, $M_z\rightarrow -M_z$, $\bm{x}\rightarrow \bm{x}+\bm{a}_2$.
\end{itemize}
In addition, for all these commensurate phases, there is another reflection symmetry tied to the topological invariance of the winding number along the $x=0$ line, that comprises a simultaneous action of $(x,y,z) \rightarrow (-x,y,z)$ and $(M_x,M_y,M_z) \rightarrow (M_x,-M_y,M_z)$; this symmetry dictates that the magnetization should lie in the $x-z$ plane on the $x=0$ line.

\section{The type of the Dzyaloshinskii-Moriya interaction}

We have considered Bloch-type DM interaction in this work and utilized a term as $\bm{M}\cdot (\nabla \times \bm{M})$ in the continuum Hamiltonian. However, we argue in this section that if the DM interaction is of N\'eel-type, our analysis and results stay valid; in particular, should we use a form like $\bm{M}\cdot [\nabla \times \left(\hat{\bm{z}} \times \bm{M} \right) ]$ in the Hamiltonian, which represents a N\'eel type DM interaction, it is straightforward to see that solutions could be obtained by rotating those discussed in the main text by $90^\circ$ around the $\hat{\bm{z}}$ axis. In other words, if a solution for the Bloch-type DM interaction is represented by $\bm{M}(\bm{r})$, a solution for the N\'eel-type DM interaction can be constructed by $\hat{\bm{z}} \times \bm{M}(\bm{r})$.

\bigskip
\bigskip
\bigskip
\bigskip

[1] Chaikin, P.M. and Lubensky, T.C., 1995. \textit{Principles of Condensed Matter Physics}, Cambridge Univ. Press, Cambridge.

\end{document}